\documentclass[journal]{vgtc}                
\ifpdf
  \pdfoutput=1\relax                   
  \usepackage{graphicx}                
  \DeclareGraphicsExtensions{.pdf,.png,.jpg,.jpeg} 
\else
  \usepackage{graphicx}                
  \DeclareGraphicsExtensions{.eps}     
\fi%

\graphicspath{{figures/}{pictures/}{images/}{./}} 

\usepackage{microtype}                 
\PassOptionsToPackage{warn}{textcomp}  
\usepackage{textcomp}                  
\usepackage{mathptmx}                  
\usepackage{times}                     
\usepackage{cite}                      
\usepackage{tabu}                      
\usepackage{booktabs}                  

\usepackage{url}
\usepackage[breaklinks]{hyperref}
\usepackage[hyphenbreaks]{breakurl}

\onlineid{0}

\vgtccategory{Research}
\vgtcpapertype{application/design study}

\title{Workgroup Mapping: Visual Analysis of Collaboration Culture}

\author{Darren Edge, Jonathan Larson, Nikolay Trandev, Neha Parikh Shah,\\ Carolyn Buractaon, Nicholas Caurvina, Nathan Evans, and Christopher M. White}
\authorfooter{
\item
 Darren Edge, Jonathan Larson, Carolyn Buractaon, Nicholas Caurvina, Nathan Evans, and Christopher White are with Microsoft Research. Email: \{daedge, jolarso, caburact, nicaurvi, naevans, chwh\}@microsoft.com
\item
 Nikolay Trandev and  Neha Parikh Shah are with Microsoft Workplace Analytics. Email: \{nitra, neshah\}@microsoft.com
}

\shortauthortitle{Edge \MakeLowercase{\textit{et al.}}: Workgroup Mapping}

\abstract{The digital transformation of work presents new opportunities to understand how informal workgroups organize around the dynamic needs of organizations, potentially in contrast to the formal, static, and idealized hierarchies depicted by org charts. We present a design study that spans multiple enabling capabilities for the visual mapping and analysis of organizational workgroups, including metrics for quantifying two dimensions of collaboration culture: the fluidity of collaborative relationships (measured using network machine learning) and the freedom with which workgroups form across organizational boundaries. These capabilities come together to create a turnkey pipeline that combines the analysis of a target organization, the generation of data graphics and statistics, and their integration in a template-based presentation that enables narrative visualization of results. Our metrics and visuals have supported hundreds of presentations to executives of major US-based and multinational organizations, while our engineering practices have created an ensemble of standalone tools with broad relevance to visualization and visual analytics. We present our work as an example of applied visual analytics research, describing the design iterations that allowed us to move from experimentation to production, as well as the perspectives of the research team and the customer-facing team at each stage in this process. 
} 


\teaser{
  \centering
  \includegraphics[width=\linewidth]{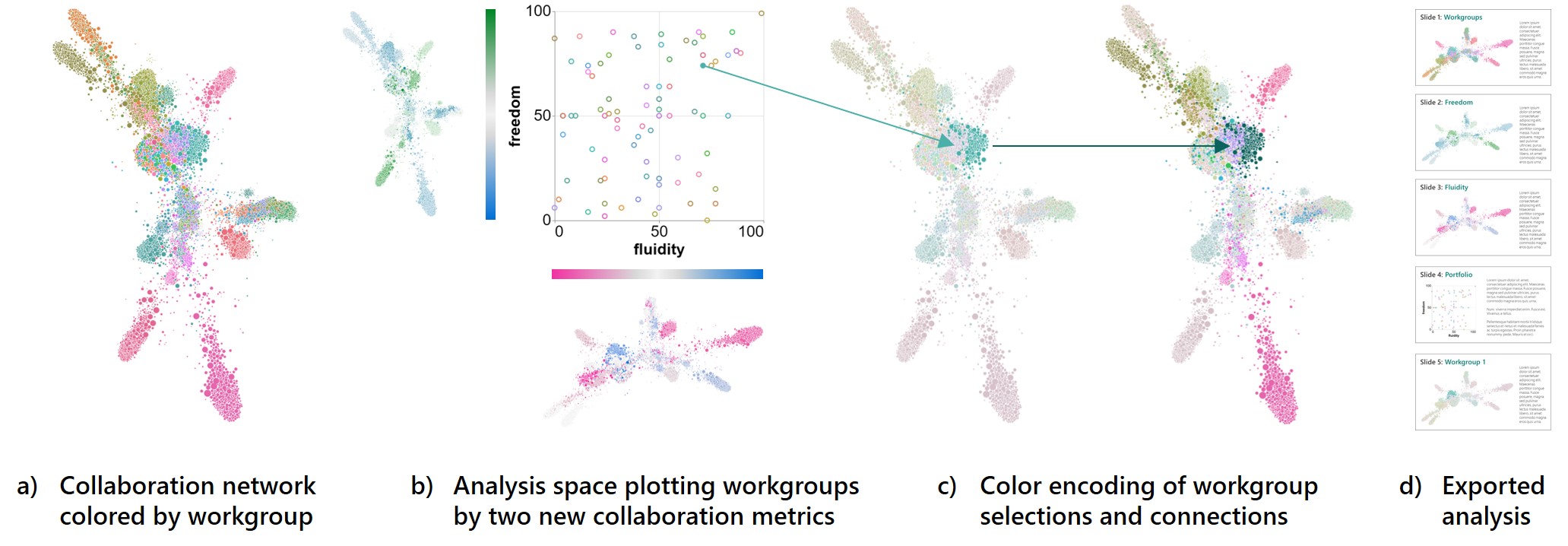}
  \caption{Visualizations of collaboration networks colored by workgroup-level collaboration metrics. An automated pipeline performs the network analysis, renders the visualizations, and produces slide presentations communicating analysis results to organization leaders.}
  \label{fig:teaser}
}

\vgtcinsertpkg

\begin{document}

\firstsection{Introduction}

\maketitle

Organizational analytics aims to advance the quantitative study of organizations using modern data, analytics, and visualization technologies. It can offer a unique lens onto the collaboration culture of an organization in terms of how work \emph{actually} gets done, which may be quite different to how organizational leaders \emph{expect} work gets done. Capturing, measuring, and communicating this gap is a crucial way to promote reflection about what kinds of collaboration culture are appropriate for different parts of the organization. We focus on the \emph{workgroup} as the unit of analysis -- informal groups of closely-collaborating individuals emerging from the dynamic needs of the organization, potentially bearing little overlap with the peer groups defined by the formal organizational hierarchy. Different kinds of workgroup collaboration are needed for different kinds of work. The approach taken in this project is one of \emph{workgroup mapping} -- creating map-like visualizations of collaboration networks that emphasize workgroup structure through both spatial clustering and nominal color encoding (Figure \ref{fig:teaser}a). Once the spatial structure is established as a frame of reference, it can support the visual analysis of multiple workgroup metrics represented using continuous color scales, as well as providing concrete grounding for more abstract mappings represented as $2 \times 2$ scatterplots (Figure \ref{fig:teaser}b). These higher-level representations reveal the overall portfolio of workgroups within an organization, supporting theories of change \cite{weiss1995nothing} regarding how shifts in collaboration culture might improve organizational outcomes. They also provide a narrative bridge to the analysis of specific workgroups based on their position in the space, again using color to guide visual attention (Figure \ref{fig:teaser}c). Since such metric-driven analysis follows a predictable structure (e.g., calling out prominent workgroups from each quadrant), we can automate a `first pass' analysis and the exporting of results to an executive-level slide presentation (Figure \ref{fig:teaser}d).  

We approached this effort as a design study in applied visual analytics research, with the goal of visualizing the structure and transformation of collaboration culture within organizations. As a collaborative project, our team includes members from Microsoft Research specializing in network analytics and visualization \cite{edge2018bringing, larson2018using, makingsenseofsearch, newsprovenance} and from the Workplace Analytics (WPA) organization responsible for both the WPA product \cite{wpa} and research and consulting on organizational analytics in general \cite{mspeopleanalytics, msinvisible, msofficemove, msemailpotential, msdatasales, msmultitask, msengaged}. Together, we formed a plan around the concept of workgroup mapping and the opportunity to test and refine this concept with executives of customer organizations who routinely meet with the WPA team for pre-scales meetings and executive briefings. The concepts and technologies developed in this paper have supported hundreds of such briefings to date -- with very positive feedback from organizational leaders -- and remain an integral part of WPA engagements with customers who elect to share their data for this purpose.

This paper makes contributions on a variety of levels. At the lowest level, it makes a system contribution to organizational analytics with impact measured through direct influence on organizational leaders. At an intermediate level, it makes multiple system and technique contributions that are of general value to visualization and visual analytics, most notably in the areas of network analysis and narrative visualization. Finally, at the highest level, it demonstrates the value of applied visual analytics research that is grounded in partners, problems, and data. Our work on such projects typically follows a common process that is evident in our published design studies \cite{edge2018bringing, larson2018using, makingsenseofsearch, newsprovenance} but not previously articulated. Here, we abstract that process into an explicit design approach, using the resulting stages to organize the incremental design iterations that characterized our development of workgroup mapping. In general, each of these iterations would not have been possible (or even conceived) were it not for the enabling capabilities provided by earlier iterations, highlighting the value of a structured design approach that systematically transitions from experimentation to production.

In the following sections, we first provide background on the related field of organizational research before expanding on the context of our design goals and the approach we took to accomplish them. We then describe each of eight design iterations, accompanied by perspectives from both the research team and the customer-facing WPA team on the progress of the work. We end with a discussion of the independent value of our technical contributions and directions for future work.

\section{Background}

Organizational analytics is closely related to the field of organizational research, which itself draws on the disciplines of sociology, psychology, organizational behavior, and management. Early developments in the field were influenced by both Max Weber's (1890s) bureaucratic model of organization \cite{waters2015weber} and Elton Mayo's (1930s) Hawthorne Studies  \cite{hawthorne1949western}, and the tension between the theory of organization and the practice of work continues to influence modern day research. The ability of organizational analytics to elucidate work practices at unprecedented levels of detail and scale means that it has great potential to bridge the gap between theory and practice, using data-driven, evidence-based models of organization. Nevertheless, its real-world application faces several challenges, including the appropriate unit of analysis, collection of data, and use of metrics to support understanding and transformation.

\subsection{The challenge of workgroup modelling}

An important object of analysis in organizational research is the \emph{workgroup}, which provides a bridge between the micro-level of individual behaviors and the macro-level of organizational statistics \cite{kirschbaum2019network}. One view of workgroups is that they are defined by formal reporting hierarchies represented visually as `org charts' or `organograms'. While the first org chart was created in 1855, it was not until the 1920s that they were adopted by business enterprises more broadly \cite{orgcharts}. Under the bureacratic model of organization, the prototypical workgroup is a group of peers reporting to the same manager within a hierarchical division of labor. In contrast, the Hawthorne Studies showed that the informal organization of work is characterized by many different kinds of relations between employees and illustrated this with early depictions of `social networks' or `sociograms' \cite{roethlisbergerwestern}. While the study of these non-hierarchical relationships helped to reveal more about the fabric of working life, the scope of such studies remained confined by formal organizational boundaries. The modern ability to collaborate freely and remotely across an organization leads to the emergence of informal workgroups and the additional challenge of determining where one workgroup ends and where others begin. This challenge can be addressed using network analytics techniques for detecting communities of tightly-interconnected nodes (e.g., Louvain \cite{blondel2008fast} or Leiden \cite{traag2019louvain}), but only if the whole network structure has been captured.

\subsection{The challenge of social network capture}

An additional challenge facing organizational sociologists in the twentieth century was the limitations of interviews and surveys as tools for \emph{sociometry}, or the measurement of social relationships \cite{moreno1951sociometry}. Using the ego-centric method of name generators and name interpreters \cite{laumann1966prestige}, each ego is first asked to list alters according to some criterion (e.g., people with whom important business decisions are discussed) before responding to a series of follow-up questions that describe each alter and contextualize the ego-alter relationship. Such questions may also probe for knowledge of alter-alter relationships as a triangulation mechanism and additional factor in calculating the strengths of relationships. All such methods are limited by the accuracy of human recall and the influence of cognitive biases (e.g., availability bias and social desirability bias). Modern technologies have significantly greater power to capture entire social networks, whether through the explicit listing of contacts on networking platforms (such as Facebook and LinkedIn), or the reconstruction of implicit networks from the logs of enterprise communication and collaboration software (such as Microsoft Office). Such approaches can also capture the dynamics of relationships over time while reducing the influence of missing data and subjectivity. However, the gain in utility to the organization from capturing, analyzing, and acting on such structures must be measured against the privacy impact for the individuals who are represented in those structures -- any metrics used to target or evaluate individuals are unlikely to be acceptable.

\subsection{The challenge of metric design}

The development of sociometric network analysis has naturally looked to the structural properties of interpersonal relationships to explain social phenomena, most notably the concept of \emph{social capital}, or the collective value inherent in social networks. There are two main schools of thought about how such value emerges: the `bridging' relationships between social groups versus the `bonding' relationships within them. The bridging school has its origins in Granovetter's study of `the strength of weak ties' -- that it is not the strong connections within groups but the weak connections between groups that allow for the diffusion of information and the influence that such bridging entails \cite{granovetter1977strength}. Burt further developed the analysis of weak ties into a theory of `structural holes' -- that advantages are conferred to individuals who span organizational networks and can broker connections between otherwise disconnected groups \cite{burt2001structural, burt2004structural}. The theory is that such brokers are better able to detect and exploit rewarding opportunities because they have earlier access to more diverse ideas. In contrast, the bonding school of social capital based on the work of Coleman \cite{coleman1990foundations} and Putnam \cite{putnam1995tuning} views network density or closure as fundamental to creating the mutual trust, norms, and sanctions that facilitate community cooperation. In an attempt to reconcile these perspectives at the team level, Burt argued that closure within teams (all pairs of individuals are connected) is necessary to fully exploit the diversity of incoming ideas that result from a lack of external constraint (external contacts are non-redundant) \cite{burt2001structural}.

Metrics based on these ideas can be applied to modern organizations (e.g., as in \cite{peopleanalytics}), but they exhibit several problems. First, they connote value judgements about individuals in ways that may create prejudice and promote bias. Second, they promote behavior change towards desirable network properties (e.g., structural holes), when simulation studies show that such universal pursuit of such proprieties dissipates their relative advantage \cite{buskens2008dynamics}. Third, they rely on teams having an unambiguous definition beyond the network structure, which is rarely the case given that team boundaries are under-specified by organizational hierarchies spanning multiple levels. Fourth, and finally, they flatten dynamic, longitudinal activity into a single and potentially misleading representation \cite{butts2009revisiting}. Overall, there is a need to develop \emph{more grounded metrics} that describe the properties of emergent workgroups, using \emph{more equitable scales} where all values have positive interpretations, based on \emph{more diverse sources} than a single network structure over a short time period. This transforms the analytic task from identifying promising (or problematic) individuals to reflecting on the fit between the portfolio of workgroups and the dynamic needs of the organization.

\section{Design Context}

\begin{figure}[tbp]
  \centering
  \includegraphics[width=\columnwidth]{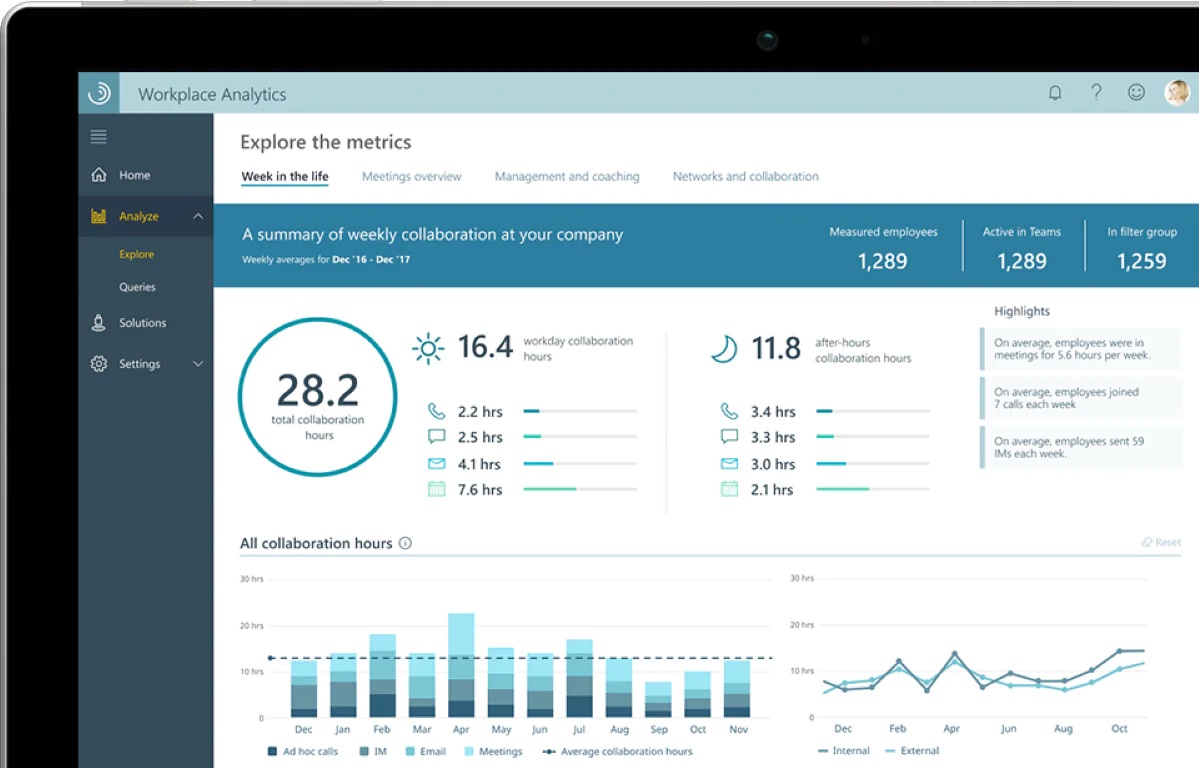}
  \caption{\label{fig:wpaapp}
  The Microsoft Workplace Analytics application showing weekly collaboration hours aggregated by month and type (IM, email, meetings). 
  }
\end{figure}

Microsoft Workplace Analytics (WPA) is an add-on application for Office 365 that aims to produce organizational insights by combining email, meeting, and instant messaging metadata from the customer organization with customer-provided business data. The results of both automated analytics and custom queries are made available through interactive dashboards that enable business leaders to view the state of their organization and monitor its transformation over time (Figure \ref{fig:wpaapp}).

Given the sensitive nature of organizational data, WPA adopts role-based access control mechanisms and EU GDPR standards on data protection and user privacy. Collaboration data from email comprises sender, recipient, and sent time, but not message content or attachments. Meetings are similarly defined by an organizer, invitees, scheduled time, attendee status, and scheduled location. The WPA product uses pseudonymized collaboration data in which user identifiers (e.g., email addresses) are replaced by cryptographically obscured strings, and only allows metrics to be reported in aggregate over a minimum sample-size threshold. These metrics include high-level summaries of weekly activities (e.g., hours of meetings vs email, internal-only vs external collaboration) combined with drill-down analyses of different facets of these activities. Examples include the amount of interaction with direct and skip-level managers and the quality of meetings based on schedule conflicts and observed multitasking (i.e., email sent while in meetings).


\subsection{Collaboration partner and research problem}

Our partner for this project was the WPA team responsible for developing innovations that anticipate future customer needs beyond the current scope of the WPA product. In preparation for executive briefings to the leaders of key customer organizations, the customer would agree to share a subset of their Office 365 tenant data with the team, and in return the team would prepare a presentation showing how new analytics could yield meaningful insights about the customer organization. Target outcomes from such meetings include customer adoption of the WPA product, consultancy on custom solution development, and feedback on ideas that may be considered for future product integration.

Our partner team wanted to use data and visualization to help organizational leaders see their organizations in a new light -- one that could lead to insights that could in turn drive positive transformation. The concept of \emph{people analytics} has been gaining currency in the business and management field as represented by the Harvard Business Review. For example, one article on \emph{Better people analytics} \cite{peopleanalytics} used the work of Ronald Burt \cite{burt2001structural, burt2004structural} to derive signatures of team-level efficiency (high internal closure, low external constraint) and innovation (low internal closure, high external constraint). Another article used sociograms and network analytics concepts to explain the findings of a 30,000 employee survey supported by hundreds of interviews -- that ``work occurs through collaboration in networks of relationships that often do not mirror formal reporting structures'' and that ``agility at the point of execution is typically created through group-level networks ... drawn throughout the organization, lateral networks across core work processes, temporary teams and task forces  ... and communities of practice that enable organizations to enjoy true benefits of scale'' \cite{agileteams}. These articles show the extent to which the tension between formal reporting hierarchies and informal workgroups remains a core concern. The WPA team has also studied how people analytics can help to drive changes in time-management \cite{mspeopleanalytics}, peer feedback \cite{msinvisible}, space planning \cite{msofficemove}, talent identification \cite{msemailpotential}, and sales development \cite{msdatasales}.

Our initial design brief was therefore to explore new methods, metrics and visualizations that would enable organizational leaders to see, reflect on, and influence the informal organization of work activity. The first opportunity for external feedback would be a presentation to the customer about their organization, using Office 365 collaboration data belonging to the customer that they had pre-approved for this purpose.


\section{Design Approach}

The design context outlined in the previous section is typical for our work in that it identifies a partner, a set of problems, and a representative dataset with which to explore potential solutions. Our goal in such collaborations is not just to develop insights from that specific dataset, but to develop interactive tools combining machine learning and machine learning that could help a wide range of people to make sense of any similar data. We first help users to view and navigate data in its original form, however, before experimenting with more advanced analytics and visualizations. This progressive approach highlights the incremental benefits of new capabilities and guides decisions about which capabilities to release for which users. Our approach to such applied visual analytics research proceeds across four stages:

\begin{enumerate}
    \item \textbf{Build Experimental Platform}. Our first step when dealing with a new problem or partner is to develop a initial system that allows us -- and our partners -- to view, navigate, and make sense of data in its initial form. This system provides both a baseline for comparison and a platform for experimentation.
    \item \textbf{Explore Candidate Capabilities}. We use the experimental platform to engage with our partners and explore the benefits of incorporating new analytics, visualizations, and interactions to the baseline system, especially those based on machine learning that have high potential levels of both risk and reward. Capabilities that show promise are incorporated into proof-of-concept systems for evaluation with partners as release candidates.
    \item \textbf{Build Production Pipeline}. Proof-of-concept systems allow us to evaluate realistic user experiences over a real dataset, but these systems need much additional work before they are production ready. We typically select one such system to build into a production pipeline that takes high-level data sources and parameters as inputs and which outputs ready-to-use data artifacts/applications.
    \item \textbf{Expand Pipeline Application}. Building a production pipeline around a single proof-of-concept requires design trade-offs towards a simpler and more focused initial experience. Once deployed, usage data and feedback from partners and end-users may suggest extensions to new data sources, analytics, or visualizations, or application to new users, use cases, or activities.
\end{enumerate}

This approach takes the Lean methodology \cite{ries2011lean} concept of releasing and continuously refining a `minimum viable product' and applies it twice -- first with partners who represent the needs of eventual end-users in the exploratory stages (1, 2), and second with actual end-users in the deployment stages (3, 4). While the approach can be used with partners who are also the target end-users, it is especially relevant when the partner owns an existing product with an established user base and is seeking to explore new product ideas with those users. In this case, direct experimentation on the deployed product risks disrupting user workflows, while direct access to users may only be possible once partner-approved proofs-of-concept are available for feedback.

Table \ref{tab:studies} shows how we have previously applied this approach with various partner teams within Microsoft: Power BI, the Digital Crimes Unit, the Bing News team, and various Bing Search teams respectively.

\begin{table}[tb]
\caption{How our previous published case studies relate to the four stages of our applied visual analytics research approach.}
\label{tab:studies}
\scriptsize%
\centering%
\begin{tabu} to \linewidth  {X[ 1 , l ] X[ 1 , l ] X[ 1 , l ] X[ 1 , l ]}

\toprule
\textbf{1. Build Experimental Platform} & \textbf{2. Explore Candidate Capabilities} & \textbf{3. Build Production Pipeline} & \textbf{4. Expand Pipeline Application} \\
\midrule
\midrule
\multicolumn4{l}{\textit{Business Intelligence: Monitoring News and Social Media} \cite{edge2018bringing}} \\
\tabuphantomline
\midrule
News corpus, text analytics, Power BI app & Custom Power BI visuals for text/metadata & Turnkey data pipeline + UI template & Expand sources to other media platforms \\
\midrule
\multicolumn4{l}{\textit{Cybercrime Intelligence: Countering Tech Support Scams} \cite{edge2018bringing, larson2018using}} \\
\tabuphantomline
\midrule
Tech support scam database, Power BI app & Analytics/vis to link/reveal related scams & Monitoring of scams by `toll free' number & Expand UI to networks of phone numbers \\
\midrule
\multicolumn4{l}{\textit{News Intelligence: Revealing Provenance of News Content} \cite{newsprovenance}} \\
\tabuphantomline
\midrule
News corpus, news reader, network vis & Text similarity techniques + configuration & Integration with Bing News index + UI & Expand users to publishers of news articles \\
\midrule
\multicolumn4{l}{\textit{Search Intelligence: Recommending Related Queries} \cite{makingsenseofsearch}} \\
\tabuphantomline
\midrule
Search query logs, model comparison app & Network modeling of related queries & Flight features for target search segments & Expand scope to other search segments \\
\bottomrule
\end{tabu}%
\end{table}

\begin{table}[tb]
\caption{Iterative development of workgroup mapping through applied research. *Indicates a generic capability that may be applied in other contexts. All of these capabilities are necessary components of our solution and would not have been developed but for the problem context.}
\label{tab:outline}
\scriptsize%
\centering%
\begin{tabu} to \linewidth  {X[ 1 , l ] X[ 1 , l ] X[ 1 , l ] }

\toprule
\textbf{Need} & \textbf{Analytics} & \textbf{Visualization} \\
\midrule
\midrule
\multicolumn3{l}{\textit{1. Build Experimental Platform}} \\
\tabuphantomline
\midrule
a. Understand the social organization of work activity & Induce collaboration networks from activity logs, e.g., for email & Create node-link diagrams (people network layout*) \\
\midrule
b. Understand the distribution of metrics across workgroups & Detect communities using Louvain and aggregate metrics & Encode the values of community metrics using node colors \\
\midrule
\multicolumn3{l}{\textit{2. Explore Candidate Capabilities}} \\
\tabuphantomline
\midrule
a. Understand the dynamics of activity within workgroups & Compute normalized structural change over time (fluidity metric*) & Represent metric evolution as node color animation \\
\midrule
b. Understand how workgroups relate to the formal org tree & Compute workgroup alignment to org tree (freedom metric*) & Plot workgroups by freedom vs fluidity in 2$\times$2 analysis space \\
\midrule
\multicolumn3{l}{\textit{3. Build Production Pipeline}} \\
\tabuphantomline
\midrule
a. Automate a generic analytics process for any target organization & Generate presentation visuals and statistics using data pipeline & Generate slides from template, visuals, and stats (SlideStamper*) \\
\midrule
b. Guide presentation and visual theming for any target organization & Explore harmonious themes from org brand color (Thematic*) & Generate harmonious visuals and slides from same theme colors \\
\midrule
\multicolumn3{l}{\textit{4. Expand Pipeline Application}} \\
\tabuphantomline
\midrule
a. Understand the implicit hierarchy of informal workgroups & Detect hierarchical community structure (hierarchical Leiden*) & Compare implicit and explicit hierarchies (future work) \\
\midrule
b. Publish illustrative analyses that do not disclose sensitive data & Generate synthetic data (network synthesis model*) & Privacy-preserving workgroup description (future work) \\
\bottomrule
\end{tabu}%
\end{table}

\section{Design Iterations}

We now proceed through each of the design iterations in which we worked with our partner team to understand the needs of organization leaders, the limitations of current solutions, and the potential for combining analytics and visualization in ways that advance the state of the art. An overview of these design iterations and how they relate to the different stages of our research approach is shown in Table \ref{tab:outline}.

\subsection{Build Experimental Platform}

The goal of presenting to the leaders of a customer organization at an in-person meeting made presentation slides an appropriate medium for delivering analytic results, with embedded images, videos, and statistics telling the story of the analysis method and its implications for the organization. The first major design choice was therefore how to represent the structure of the organization and its workgroups.

\label{orgvis}
\subsubsection{Organization visualization}

Much visualization research has focused on the visual representation of tree structures, which are the foundation of formal reporting hierarchies. This includes space-filling TreeMaps \cite{shneiderman1992tree}, hybrid TimeTrees \cite{card2006time}, comparative CandidTrees \cite{lee2007candidtree}, and ``organic org charts'' whose animations show the evolution of organizational structure \cite{orgorgcharts}. Research projects have also explored embedding lateral network  connections within an `expand on demand' tree structure of hierarchical network connections, using a branching tree repesentation in TreePlus \cite{lee2006treeplus} and a radial tree representation in TreeNetViz  \cite{gou2011treenetviz}. For our use case, however, our visualizations needed to stand alone, the identities of individuals were not important, and the unknown structure of collaboration took precedence over the known structure of the organization. A network-oriented base representation was therefore desirable. Despite the idea that node-link network representations `do not scale' as effectively as adjacency matrices \cite{ghoniem2004comparison}, we observe that this relates to small (20--100 node) networks rather that the much larger networks of organizations. We take more inspiration from work on \emph{graphs as maps} \cite{gansner2010gmap,jianu2014display}, where the `synthetic geography' of the graph supports the cartographic tasks of orientation, wayfinding, and exploration by region. We therefore used force-directed layout of unlabelled node-link diagrams as our basic visual notation, with people nodes and collaboration links.

From the collaboration data available to us, we created a network edge for each pair of individuals that exchanged at least 4 emails within a month time period (with at least 1 email in each direction), and focused our analysis on the largest connected component. For such people networks, we typically observe modularity scores (how well networks can be clustered into coherent `communities' of nodes \cite{newman2004finding}) in the region 0.6--0.8, representing a strong community structure.

Of the standard algorithms for force-directed network layout, ForceAtlas2 (FA2) \cite{jacomy2014forceatlas2} does a particularly good job of spatializing the apparent community structure. However, FA2 performs poorly when applied to random network layouts on the order of 10,000--100,000 nodes, which is typical for the kinds of organizations we are targeting with this work. Simulated annealing layout algorithms like OpenOrd \cite{martin2011openord} do not have such scaling issues and may be used to create an appropriate initial layout for FA2, resulting in good performance and clear community clustering. In general, no one algorithm achieves acceptable layouts on all network types in a fully-automated way. For organizational networks, we performed many experiments using different parameters and algorithms in the open-source Gephi application \cite{bastian2009gephi} that is ideal for exploratory network visualization. The final `people network' layout process we created and used for workgroup mapping is as follows:

\begin{enumerate}
    \item Apply the liquid and expansion phases of OpenOrd to separate proto-communities into their own region of the plotting region.
    \item Apply FA2 expansion (high scaling parameter) to ensure that there is enough space to render all nodes within the overall network.
    \item Apply FA2 contraction (low scaling parameter with no-overlap and strong gravity), creating a compact layout with clear structure.
\end{enumerate}

\subsubsection{Workgroup metric visualization}
\label{metricvis}

The Louvain algorithm \cite{blondel2008fast} is the standard approach to network community detection that aims to maximize the modularity \cite{newman2004finding} of the community partition. We first used the Louvain implementation built into Gephi to detect node communities, before encoding community at the node level using colors drawn from a Gephi color palette. The resulting color distribution reinforces the spatial partitioning of communities already present in the layout and enables the viewer to evaluate the quality of the community detection. Any node-level metric may also be aggregated to the community level and mapped to node color using a continuous color scale. Such community-level aggregation hides the contributions of individuals in a way that preserves privacy. Beyond community, the only node-level information we preserve in our layouts is node degree, which we map to node size to create visual landmarks for orientation. Given the number and diminished importance of individual links in such dense network layouts, we also tend to hide the links to create node-only workgroup maps. An example layout and set of community-level node color encodings is shown in Figure \ref{fig:teaser}. Our subsequent use of the term `workgroup' refers to these network-theoretic communities derived from collaboration networks.

\subsubsection{Experimental Platform Discussion}

The senior executives we work with have deep familiarity with the official organizational hierarchy in their company.  In most cases, they have limited experience with organizational analytics and network modeling.  To enable our executive audience to understand these new concepts quickly without being overwhelmed by technical details, we aimed to create a resonant storyline and present insights about their company within the context of that story. At the end of the presentation, we wanted the audience to (1) see their organization through a new network lens, (2) understand how this lens can be applied to notions alike agility and innovation, and (3) start asking important questions that can be answered through further analysis using these new methods.

We thus see workgroup mapping as an introduction to the use of data, networks, and visualization for organizational sensemaking, allowing executives to look beyond traditional business metrics and grasp -- perhaps for the first time -- the existence and importance of an implicit organization defined by informal collaboration. Only by seeing such a structure in concrete terms can executives enter a feedback loop in which their visual insights lead to the development and roll-out of new policies and programs aiming to influence collaboration culture. Executives often expressed surprise at the views of their organization we presented to them, and a desire to use their new understanding to drive positive change. Our visualizations similarly helped executives to realize that they can take a scientific approach to understanding complex initiatives like digital transformation by measuring the properties of the collaboration structure. The unique glyph of a company also allowed us to compare it to other companies and industries, as well as illustrating concepts like `siloed' or `connected' in clear visual terms.

\subsection{Explore Candidate Capabilities}

The kinds of capability we were exploring in this project were metrics that could be computed and visualized at the workgroup level. Using our experimental platform in Gephi, we were able to illustrate the structural distribution of a wide variety of network metrics, including degree centrality, betweenness centrality, density (closure), clustering coefficient, and constraint. While each was useful in its own way, none shed light on the two most pressing concerns from both the WPA team and our literature review -- network dynamics over time as they relate to organizational agility \cite{butts2009revisiting} and the relationship between informal workgroups and formal reporting structures \cite{agileteams}. We now present the two new metrics that we developed in response to these concerns.

\subsubsection{Fluidity metric}

For our visualization of organizational collaboration, we constructed a longitudinal collaboration network from strong email exchange relationships over the whole period of data collection. This provided a single, stable frame of reference for the display and comparison of multiple metrics. It also allowed us to illustrate the dynamic variation in metrics over time, by exporting videos that communicated changes in metric values as animated changes in node colors. A limitation of this approach, however, and a critique of network modelling in general \cite{butts2009revisiting}, is that such mono-structures do not capture changes over time.

If a longitudinal workgroup exhibited high levels of structural change, it may be associated with the positive qualities of agility, brokerage exploitation, and organizational learning. Similarly, low levels of connectivity change may be associated with the reliability of communication channels and the efficiency of organizational execution.

Many visualization techniques support the representation and interactive exploration of dynamic networks, using animation, sequencing, juxtaposition, and integration to encode change over time \cite{beck2017taxonomy}. Almost all techniques require a time series of graphs as input, which introduces the challenge of specifying in advance the correct size and overlap of a sliding time window to capture phenomena of potential interest. The DynNoSlice system \cite{simonetto2017drawing} does away with this constraint, however, by performing force-directed layout on node `noodles' in a space-time cube. The relative positions and movements of nodes in any given timeslice are therefore determined by relative connectivity and changes in connectivity respectively. We wanted a collaboration metric that could capture such qualitative structural changes in quantitative form.

There are two principled methods for quantifying statistical changes in dynamic network structure. The first, \emph{scan statistics} \cite{priebe2005scan}, detects anomalies as ``local regions of excessive activity'', where the region scale can be progressively expanded from individual nodes to include their $N^{th}$-degree connections. The second, \emph{omnibus embedding} \cite{levin2017central}, uses network machine learning (spectral embedding) to jointly embed a time series of networks into a single metric space where node distances may be interpreted as `relatedness'. Since each node is then represented by a time series of positions in the embedded space, the distances between these positions captures `change in connectivity' as it impacts the latent structure. This means that it performs \emph{normalization} -- mitigating any bias from node degree or degree distribution -- while also allowing \emph{aggregation} across arbitrary node sets (e.g., workgroups extracted as communities from the longitudinal collaboration network). For these reasons, we use omnibus embedding as the foundation of a new `fluidity' metric for collaboration networks (Figure \ref{fig:fluidity}), as follows:

\begin{figure}[tbp]
  \centering
  \includegraphics[width=\columnwidth]{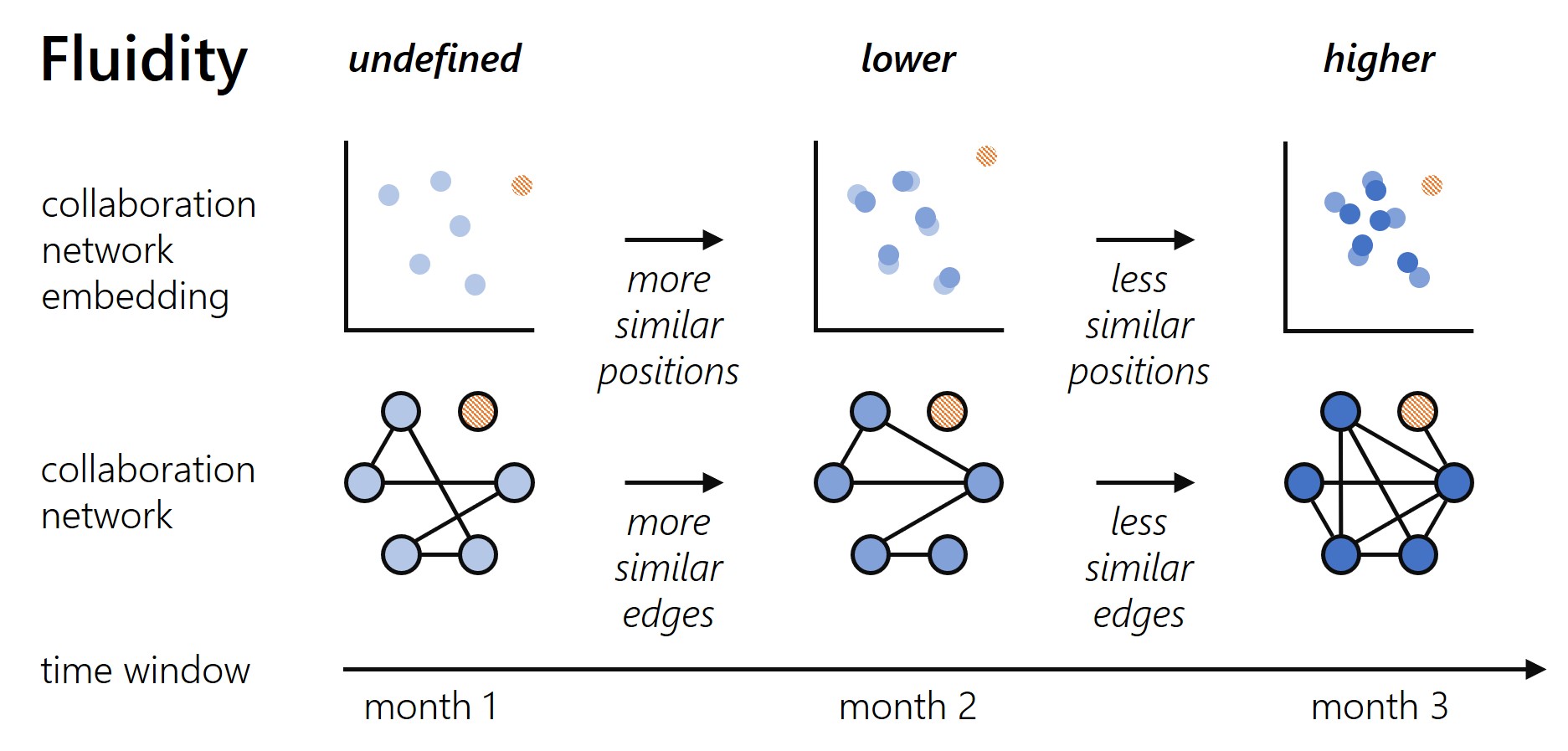}
  \caption{\label{fig:fluidity}
  How fluidity quantifies the variability of collaboration over time.
  }
\end{figure}

\begin{enumerate}
    \item Create an omnibus embedding for each adjacent pair of monthly collaboration networks, defining the embedded space using the most explanatory eigenvectors of the omnibus adjacency matrix.
    \item For each embedding and for each individual, use the cosine similarity between their nodes across time windows to represent that individual's change in connectivity month-over-month.
    \item Compute \emph{individual fluidity} scores as the mean of the month-over-month changes for that user and \emph{workgroup fluidity} scores as the mean fluidity for all individuals in the workgroup. 
\end{enumerate}

The overall fluidity of a workgroup $w$ over time ($t=0..N)$, where each member of $w$ is represented by a position vector $i$ in the omnibus embedding of successive collaboration networks, is thus as follows:

\begin{equation}
    Fluidity_{w} = \sum\limits_{t=1..N}(1 - \sum\limits_{i\in w}cosine\_similarity(i_t,i_{t-1})/|w|)/N
\end{equation}

Note that individual fluidity scores are unimportant here beyond their contribution to workgroup fluidity. While an individual's fluidity may be determined by many forces beyond their control, strong signals at the workgroup level indicate the presence of a dominant force that we may ascribe to the underlying \emph{culture} of that workgroup. The distribution of workgroups by fluidity thus provides a lens onto the \emph{dynamics} of collaboration within an organization. Changes in these dynamics over time may be revealed through the use of a rolling time window over the underlying collaboration data. While shorter time windows allow changes to reveal themselves sooner, time windows that are too short might not accurately capture typical workgroup behavior.

Omnibus embedding is available through two open-source Python packages: GraSPy for graph statistics \cite{graspy} and the topologic package (used in our pipeline) for network induction and analytics \cite{topologic}.

\subsubsection{Freedom metric}
\label{sec:freedom}

To measure the differences between formal reporting hierarchies and informal collaboration networks, we developed a new metric from first principles. For a workgroup with perfect alignment: (1) the minimum spanning tree (MST) connecting all members of a workgroup across the formal reporting hierarchy would consist entirely of workgroup members (non-division of reporting lines), and (2) apart from the root node, all peers of all MST nodes in the formal reporting hierarchy would be members of the workgroup (non-division of peer groups except for team leaders). We used these criteria to create a continuous alignment scale -- penalizing the division of reporting lines and peer groups respectively. We refer to the complement of alignment as `freedom', as it measures the expressed freedom of workgroups to form across organizational boundaries (Figure \ref{fig:freedom}). For a workgroup $w$:

\begin{figure}[tbp]
  \centering
  \includegraphics[width=.9\columnwidth]{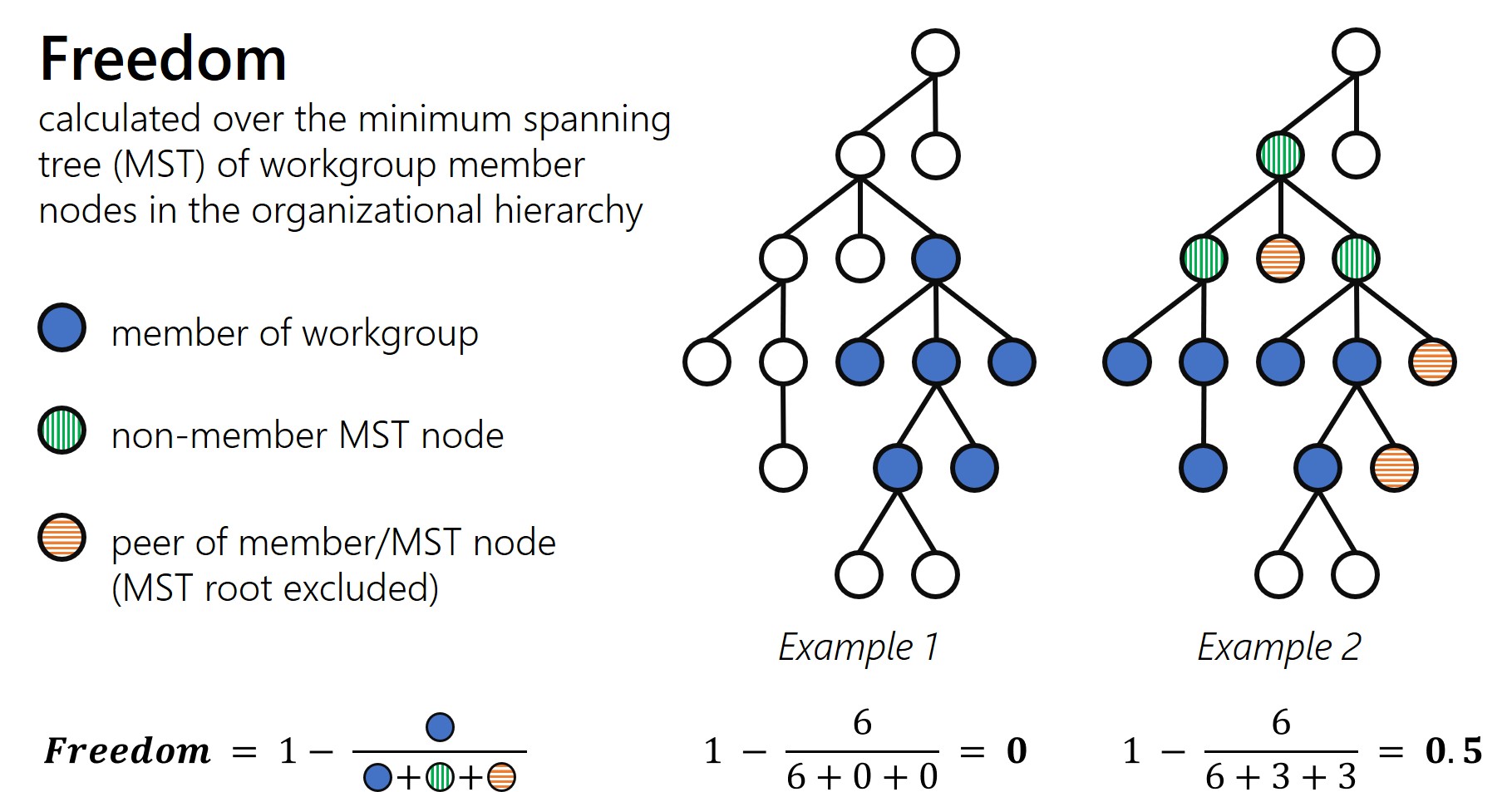}
  \caption{\label{fig:freedom}
  How freedom quantifies hierarchy-spanning collaborations. 
  }
\end{figure}

\begin{equation}
    Freedom_w = 1 - \frac{|\:w\:|}{|\:MST_w\ \cup \ peers\ of\ non\mbox{-}root\ nodes\ in\ MST_w\:|}
\end{equation}

Given that reporting hierarchies may change over time, we compute the freedom of a workgroup based on the reporting hierarchy at the middle of each month and take the mean across all months of analysis. 

As with fluidity, there are positives and negatives associated with all levels of freedom. High freedom workgroups are important as the `connective tissue' of the organization, whether expressed as lateral networks, temporary teams, task forces, or communities of practice \cite{agileteams}. At the same time, such workgroups can face challenges in terms of visibility, accountability, and responsibility. Many organizations would also fail to function in an effective and predictable way without a foundation of workgroups closely aligned with the reporting hierarchy.

\subsubsection{Candidate Capabilities Discussion}

It was at this stage that we began developing a story about visualizing organizational culture through the lens of workgroup freedom and fluidity. 
We needed to create presentation slides that communicated both the method in general and its application to a specific organization. We framed this as a design challenge requiring \emph{narrative visualization}.

While the concept of narrative visualization emerged from studies of journalistic storytelling \cite{segel2010narrative}, subsequent work has examined the impact of sequence on data storytelling in the linear format of slide presentations \cite{hullman2013deeper}. Drawing on this work, our presentation structure adopted the concepts of \emph{parallelism} and \emph{measure walks}, in which there is a transition from a general structure (the whole network colored by workgroup) to a specific measure encoded over that structure (freedom), with multiple notable substructures (workgroups) discussed in sequence in the context of this measure, before returning to the general structure and repeating for the next measure of interest (fluidity).

\begin{figure}[tbp]
  \centering
  \includegraphics[width=\columnwidth]{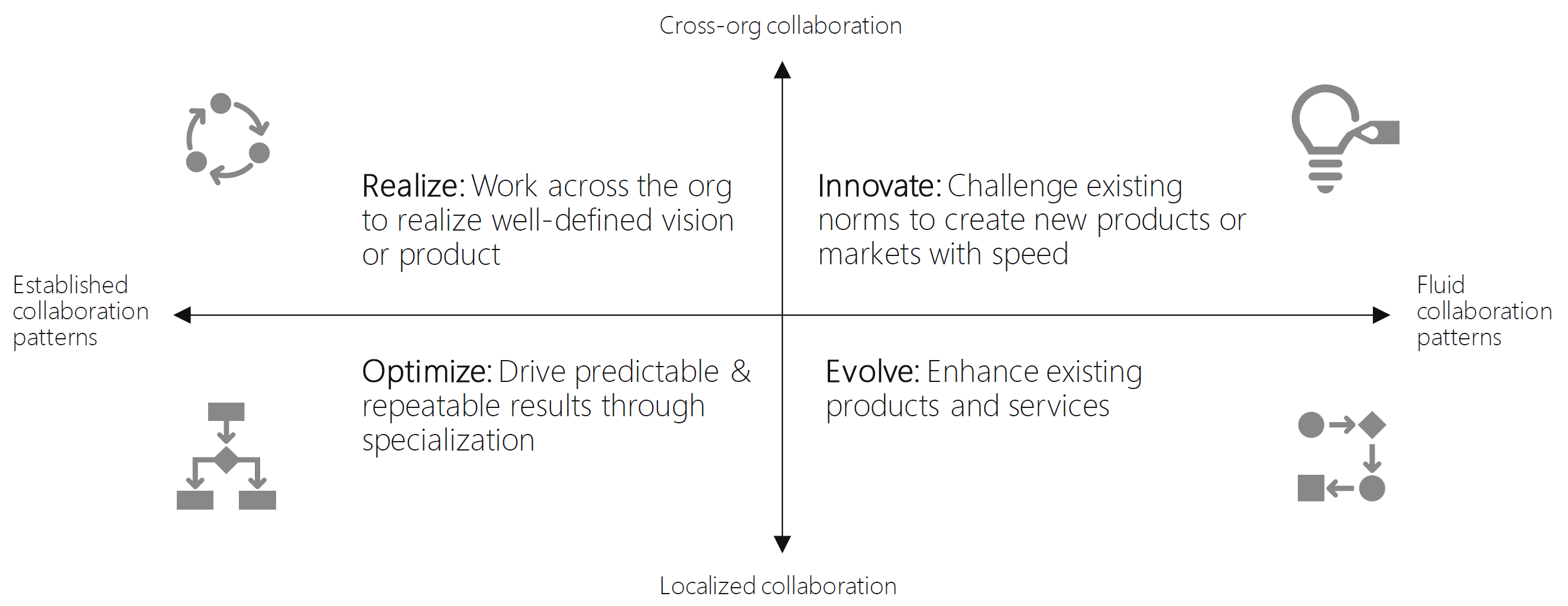}
  \caption{\label{fig:quad}
  Mapping of freedom and fluidity metrics into a $2 \times 2$ analysis space. The freedom axis runs from localized collaboration (low) to cross-org collaboration (high), while the fluidity axis runs from established collaboration patterns (low) to fluid collaboration patterns (high). Workgroups may be plotted based on their individual metric values, while organizations (or even industry verticals) may be plotted based on average metrics.
  }
\end{figure}

The freedom and fluidity metrics allow us to quickly represent highly complex and layered notions within organizations, like the agility and effectiveness of reporting structure design. Since we were concerned that our audience would not have the time for complex scientific explanations, we relied on diagrammatic illustrations to quickly communicate these concepts without an explanation of mathematical definitions. These became a standard, easily understood part of the presentation. Similarly, visualizing the metrics as a $2 \times 2$ analysis space (Figures \ref{fig:teaser}, \ref{fig:quad}) allowed us to use a familiar technique commonly used to evaluate and think about teams or products within their businesses (e.g., \cite{burt2001structural}). The visualization has generated many interesting conversations around how a certain distribution can be explained by existing cultural norms, or how strategic intentions may diverge from actual collaborative patterns. Employing a simple summary visualization also resonated with our goal of representing the organization uniquely, yet easily allowing the executive to understand workgroup composition.


For an initial series of 5 customer presentations, the WPA team would prepare data sharing agreements and share data with the MSR team. The MSR team would then generate all necessary images and statistics, manually integrate these into a PowerPoint deck, and return to the WPA team for presentation to the customer. These occurred at a rate of 1 presentation every 2-4 weeks, with the framing, structure, and language all evolving with each round of customer engagement and feedback. Although each iteration resulted in a significantly leaner and more focused presentation, assembling it was still a painstaking and labor-intensive process. A compounding disadvantage was that such manual production costs create resistance to change in what should otherwise be an exploratory process. It was therefore time to build a production pipeline that could scale to many customer engagements.

\subsection{Build Production Pipeline}

We developed our production pipeline as an Azure App Service \cite{appservice} using Python, with an initial focus on automating the induction of networks from collaboration logs, the detection of `workgroup' communities, the computation of our freedom and fluidity metrics, and the generation of data graphics and statistics for presentation. This work involved taking the human out-of-the-loop at several stages of our experimental workflow. Rather than detecting communities `manually' by running Louvain inside Gephi (Section \ref{metricvis}), we used the Python-Louvain package \cite{python-louvain} with NetworkX. Similarly, rather than running our people network layout process across several manual stages in Gephi, we automated the process using Gephi Toolkit in Java \cite{gephi-toolkit} (used to create the layout in Figure \ref{fig:teaser}). Once these stages of the pipeline were completed, the major outstanding task was to automate the assembly of data graphics and statistics into a final slide presentation. 

\subsubsection{SlideStamper}

To create a truly scalable pipeline, we needed to find an effective way to automate the production of presentation slides with minimal human intervention. Ideally, such automation would also allow us to decouple our iterative development of the general presentation narrative from our analysis of any specific customer data, such that new analytics or narrative structures could be retrospectively applied to any existing analysis, while reusing all pre-computed data graphics and statistics.

Programmatic generation of PowerPoint presentations is supported by the Open XML SDK 2.5 \cite{openxml}, which enables strongly-typed generation of the XML markup that comprises the Microsoft Office document formats. In related work, we have previously used this SDK in the HyperSlides system \cite{edge2013hyperslides} to convert hierarchical presentation outlines into networks of hyperlinked slides for non-linear delivery. We have also developed prior systems to generate hybrid slide-canvas media from such outlines \cite{edge2016slidespace}, as well as presentation slides from established narrative templates \cite{pschetz2014turningpoint}. Here, we use the SDK within an Azure Function \cite{azurefunctions} to create a web service called \emph{SlideStamper} that assembles PowerPoint presentations on demand from several inputs:

\begin{enumerate}
    \item Media in standard image or video formats (e.g., data graphics generated by our pipeline), provided either as local files or urls.
    \item A PowerPoint template with object placeholders for media and text placeholders for statistics and other custom text. Placeholders tags are indicated by user-defined strings, e.g., ``\{global network visualization\}'' and ``\{detected workgroup count\}''.
    \item A JSON specification describing: a) how to replace placeholder tags with uploaded/linked media or text, including an optional url for interactive hyperlinking; b) how to reuse slides and slide sequences with alternative replacements (e.g., for different workgroups and metrics); and c) the color scheme for the final slides.
\end{enumerate}

\begin{figure}[tbp]
  \centering
  \includegraphics[width=.8\columnwidth]{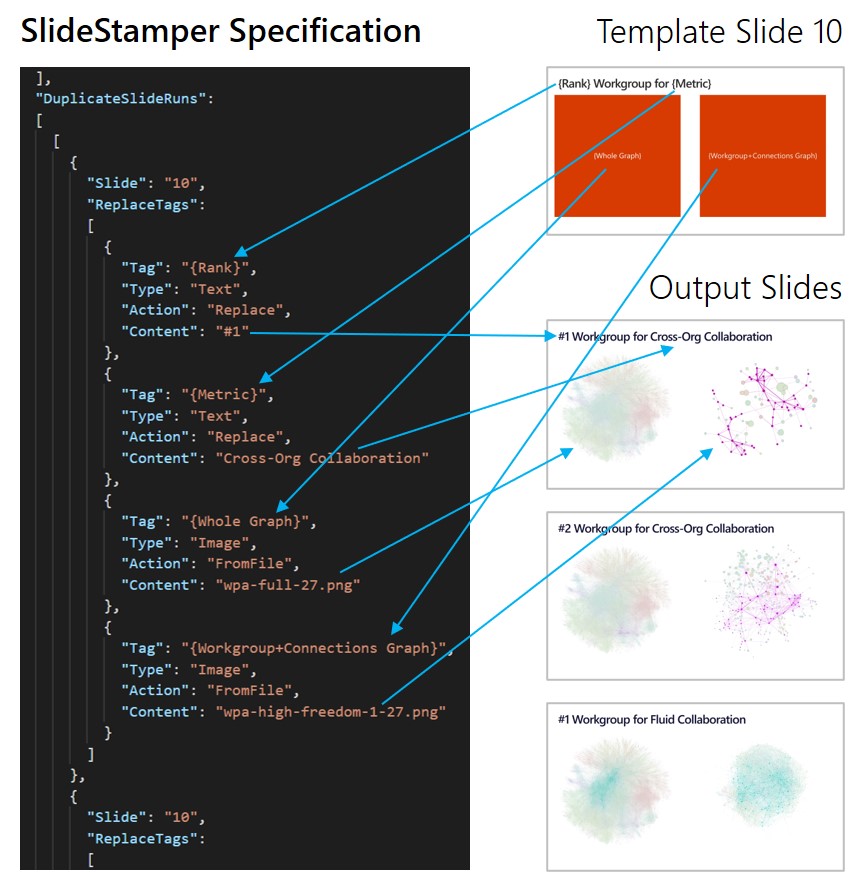}
  \caption{\label{fig:slidestamper}
  SlideStamper. A web service that generates PowerPoint presentations from a JSON specification of the final presentation color scheme and how to replace template placeholder tags with final text and images.
  }
\end{figure}

An example of how the JSON specification enables template slides to be reused in multiple presentation sequences is shown in Figure \ref{fig:slidestamper}. 

This approach aligns with the natural workflow of exploratory design, allowing a concrete prototype presentation to be built first before abstracting the emergent design into a reusable template. While the same narrative structure can be applied to any customer organization, it is also the case that color schemes generally need to be created on a per-organization basis to reflect brand colors and visual style. For major customer engagements, this task was performed by graphic designers who would research the house style of the customer organization. However, a color scheme designed for a presentation does not necessarily work when applied to networks and other visualizations, and vice versa. There are also overheads associated with ensuring color consistency across data graphics and their presentation context. To make the approach more scalable, we needed to make this process as close to automated as possible, while at the same time retaining human input into what can be a highly subjective process. 

\subsubsection{Thematic}

In response to the challenges of creating consistent color schemes spanning data graphics and presentations, with specific support for the color requirements of network visualizations, we present \emph{Thematic} -- a collection of tools comprising an interactive web application developed in React \cite{react}, a JavaScript library, and a Python package. Thematic draws on a rich history of color research for information visualization, most notably the perceptually-graded cartographic color scales pioneered in ColorBrewer \cite{harrower2003colorbrewer}. Recent research has explored how to automate the production of such color scales in ways that reflect real design practice \cite{smart2019color}, how to explore a space a color schemes using direct manipulation \cite{shugrina2019color}, and how to automate the color selection process for both multivariate scatterplots \cite{wang2018optimizing} and continuous fields \cite{cheng2018colormap}.

Thematic is different from existing tools in that it supports the simultaneous creation of a presentation color scheme (including foreground text and accent colors and a slide background color) and color scales designed to operate within this context. Also unlike existing color tools, it supports the three levels of nominal colors that are often required for network visualization: a standard color (e.g., to show community membership and structure), a bold color (e.g., to highlight a community), and a muted color (e.g., to show network context). In practice, this requires both the lightness and saturation color dimensions to be reserved for scale differentiation, with hue as the primary means to differentiate adjacent nominal categories (e.g., node communities, which in our case are organizational workgroups). Sequential and diverging color scales are provided along with these nominal scales (e.g., for showing continuous or quantized metric values), with the base lightness of all scales chosen in a way that provides sufficient contrast with the slide background (which may be a tinted off-white or off-black color). In line with modern presentation and user interface design, the color schemes are also invertable to support light and dark modes.

To support this functionality, we use the HSLuv \cite{hsluv} color space and library. HSLuv is a cylindrical projection of CIELUV in which angle denotes hue, depth denotes lightness, and radius denotes the proportion of available chroma (absolute level of perceived color) for the given hue and lightness. HSLuv thus provides perceptual uniformity in the hue and lightness dimensions and guarantees all saturation values produce valid colors. We also use the chroma-js implementation of HCL \cite{chromajs} to equalize absolute chroma levels across different background hues.

Specification of a color scheme is performed using a set of six sliders manipulating different aspects of the overall color relationships:

\begin{enumerate}
    \item \textbf{Accent hue}. HSLuv input selected on a scale of 0--360, or derived from a seed color selected via a standard color picker.
    \item \textbf{Accent saturation}. HSLuv input selected on a scale of 0--100.
    \item \textbf{Accent lightness}. HSLuv input selected on a scale of 0--100. Values in the range 40-60 best support light/dark mode inversion.
    \item \textbf{Background level}. Theme input selected on a compound scale of 0--100 defining background contrast and style. 0--50 interpolates the background from grayscale to a maximum background saturation $s_{max}$ for a fixed lightness ($l=95$ for light mode, $l=10$ for dark mode), while 50--100 interpolates the background lightness to white/black while maintaining $s_{max}$. For a given hue and lightness, $s_{max}$ is the maximum saturation that results in a colors whose chroma does not exceed 4 (giving equally colorful but muted colors suitable for slide and user interface backgrounds).
    \item \textbf{Background hue shift}. Theme input selected on a compound scale of 0--100 defining the relationship between the accent hue and the background hue, using the concepts of analogous and complementary hues from color theory. Values in the range 50--100 specify clockwise hue shifts, with the subrange 50--75 allocated to analogous colors ($+[0,30]$) and  75--100 to complementary colors ($+[150,180]$). Similarly, values in the range 50--0 specify counter-clockwise hue shifts, with the subrange 50--25 allocated to analogous colors ($-[0,30]$) and  25--0 to complementary colors ($-[150,180]$). All shifted hue values are taken modulo 360.
    \item \textbf{Nominal scale step}. Theme input selected on a compound scale of 0--21 defining the direction and spacing of nominal hue colors. Values in the range 11--21 specify clockwise hue shifts, while values in the range 0--10 specify counter-clockwise hue shifts. The next hue is selected as the closest hue in the direction of traversal that is maximally-different to all hues selected thus far. Values of 10 and 11 take 3 equally spaced hues from the first traversal of the color wheel, 9 and 12 take 4 equally spaced hues, and so on up to 0 and 21, which take 13 hues respectively. The first two nominal hues also form the basis of sequential and diverging color scales, with the opposing colors of diverging scales set 90 degrees apart to avoid pairs of hues affected by color blindness.
\end{enumerate}

\begin{figure}[tbp]
  \centering
  \includegraphics[width=\columnwidth]{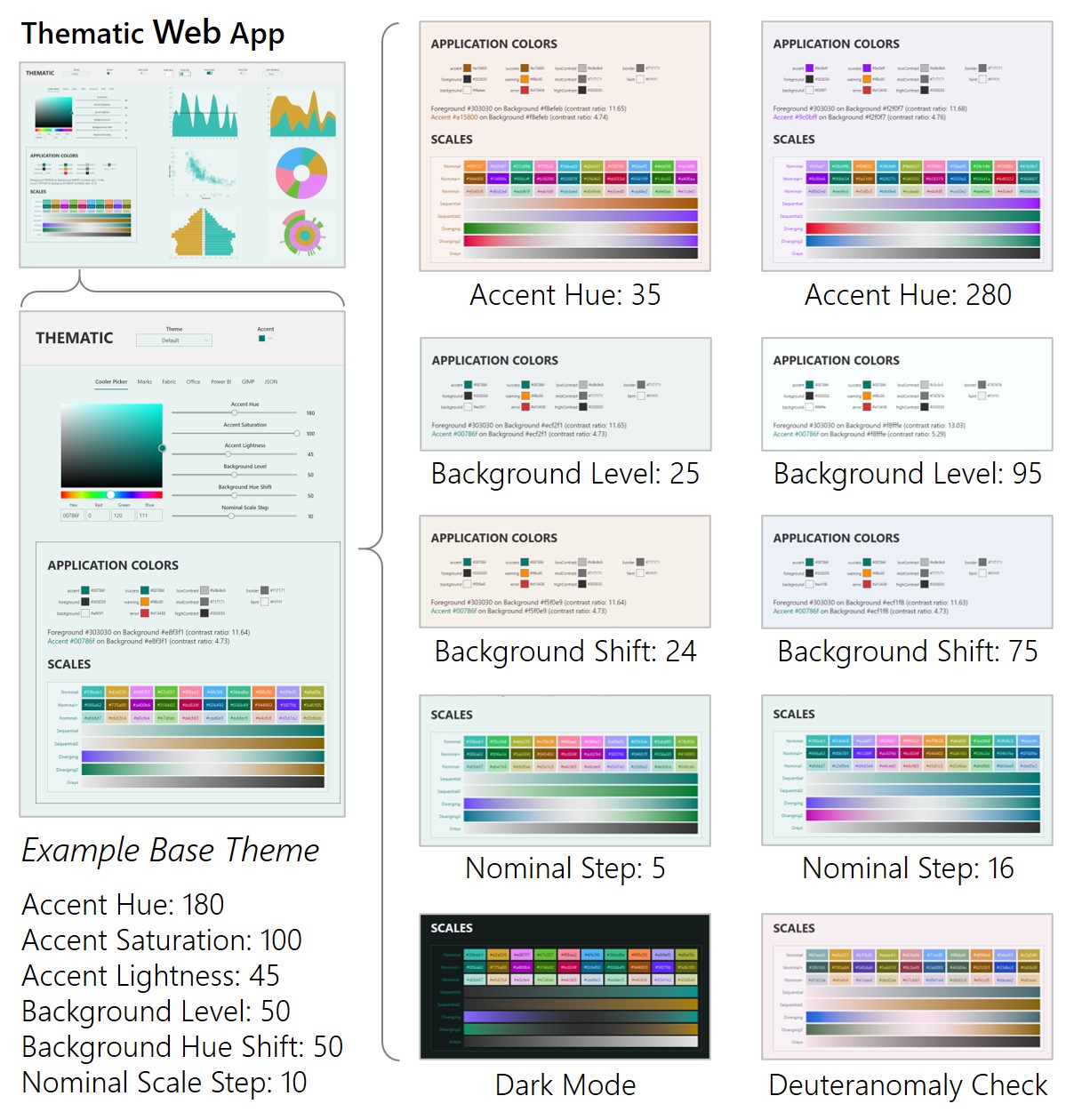}
  \caption{\label{fig:thematic}
  Thematic. A web application that enables the visual construction of themes spanning color scales, visualization marks, and application chrome. Color scales optimized for network visualization are explored using a set of six sliders that enforce harmonious color relationships.
  }
\end{figure}

The Thematic interface and several example color schemes are shown in Figure \ref{fig:thematic}, while its application to network visualization is shown in Figure \ref{fig:teaser}. Beyond the color scheme controls described above, Thematic provides automated checks for contrast ratios \cite{contrast-ratio} and color-blindness \cite{color-blind}. It also allows specification of chart components (e.g., areas, axes, marks) and other interface elements designed to be read by downstream generators of data graphics, and exports application-specific theme files for Microsoft Office, Microsoft Power BI, and GIMP, along with a Microsoft Office Fabric theme and a generic JSON specification. Thematic is itself styled by the specification under construction, enabling exploratory evaluation of color schemes in context.

\subsubsection{Production Pipeline Discussion}

The pipeline as described allowed the WPA team to significantly increase the speed at which they could experiment with new narrative structures and create customer-ready presentations.  It also allowed them to fully scale the capability across the entire global salesforce of 30+ dedicated salespeople for Workplace Analytics. This became a core component of the sales process, with demand rapidly growing to surpass almost all other pre-sales assets in both interest and sales impact. From a baseline of 1 presentation every 2--4 weeks with the manual process and 5 presentations created in this manner, the WPA team were able to use our pipeline to create and deliver up to 4 customer presentations per week. At the time of writing, our pipeline has been used many hundreds of times, helping us to reach and influence a broad cross-section of leaders of major organizations.

\subsection{Expand Pipeline Application}

In this section, we describe two additional capabilities that aimed to increase the success of the production pipeline -- first by revisiting the nature of workgroup detection in light of customer feedback, and second by developing a way to share details of our approach without disclosing sensitive information about any particular organization.

\subsubsection{Hierarchical workgroup detection}

The Louvain community detection algorithm \cite{blondel2008fast} was a fundamental part of our pipeline, not just for detecting informal workgroups but as a means of \emph{defining} them as emerging from the organic structure of interpersonal collaboration. At the same time, being able to detect and analyze \emph{hierarchies} of workgroups spanning multiple scales of organization would give customers additional control over the scale of interest for any given analysis. However, the Louvain algorithm returns a single optimal partition rather than a hierarchy of parations, and is known to have a \emph{resolution limit} whereby it fails to identify modules below a particular size \cite{fortunato2007resolution}. The achievable resolution ultimately depends on the overall network size and the degree of connectedness between communities, but may be significantly larger than desired.

In the search for an improved approach, the constant Potts model (CPM) \cite{traag2011narrow} aims to maximize the number of within-community edges while maintaining a relatively small number of communities. This approach has been proven to be `resolution-limit--free' in that the partitioning is invariant over all induced subgraphs. A resolution parameter is used to set the threshold between intra- and inter-community density, with higher resolutions leading to the detection of more communities.

However, the application of Louvain optimization to the CPM metric still has the potential to produce communities that are arbitrarily badly connected. The recent development of the Leiden algorithm \cite{traag2019louvain} counteracts this problem and can be used to maximize either modularity or CPM, with connectedness between communities that is guaranteed near-optimal. However, while the sizes of detected communities can be varied by manipulating a resolution parameter, there is no universal mapping between resolution values and community sizes. To detect communities of a target size, we therefore adopt an iterative approach with a fixed resolution (1.0) in which at level $N$ in the community hierarchy, we apply Leiden to the induced subgraphs of communities detected at level $N-1$. This process terminates whenever all detected communities are below a pre-specified maximum community size. Since the Leiden algorithm is efficient (two orders of magnitide faster than Louvain for large networks), we are able to run our approach many times with varying maximum community sizes, monitoring the effects on overall modularity. We observe a general trade-off between smaller maximum community sizes and higher modularity scores, and use an elbow-finding routine over the community-size--modularity curve to suggest a maximum community size that balances this trade-off, and use this size as the starting point for multiscale workgroup mapping.


\subsubsection{Hierarchical workgroup synthesis}

A major challenge of communicating organizational analytics methods is that the derived artifacts are often tied to sensitive data belonging to an organization. Obtaining permission to share such artifacts is challenging from the perspectives of both individual privacy and organizational secrecy. Our approach to sharing details of workgroup mapping independently of customer data has been to generate synthetic networks and data that embody similar statistical and structural properties to actual data but without the disclosure risks.

A variety of generative models for random networks have been developed, beginning with the Erdős--Rényi model \cite{erdHos1960evolution} in which each edge has an identical probability of being present. The Watts--Strogatz model \cite{watts1998collective} aimed to improve on such fully random graphs by randomly rewiring a regular lattice structure, facilitating the emergence of locally-clustered regions with short average path lengths -- key qualities of `small world' networks commonly observed in the real world. One significant artifact of this model, however, is an unrealistic degree distribution that does not capture the `scale free' distribution typical of many real networks. The Barabási--Albert model \cite{barabasi1999emergence} does achieve such a degree distribution through a `preferential attachment' mechanism in which nodes that are higher in degree have a higher probability of receiving new connections. The downsides of this model are that it doesn't lead to local clustering (i.e., the formation of closed triangles) or global community structure. The Lancichinetti–Fortunato–Radicchi benchmark \cite{lancichinetti2008benchmark} addresses many of these challenges by sampling degree distributions and community size distributions from truncated power law functions and using a single global parameter to adjust the level of community mixing. The stochastic blockmodel \cite{wang1987stochastic} and related family of techniques aim to capture variations in the community-like `blocking' observed in social networks, achieved by explicitly defining the likelihood of connections within and between blocks defined over a network adjacency matrix. However, such blockmodels can fail to generate the desirable small-world and scale-free properties.

We thus adopt a hybrid approach to  synthesizing collaboration networks, with all sampling from an a representative power-law distribution: (1) sample community sizes for a target number of communities; (2) within each community, use preferential attachment with sampled edge weights to create local hubs; (3) between communities, use preferential attachment with sampled edge weights to create global hubs and small average path lengths; (4) repeat steps 1--3 as needed to create and connect a hierarchical community structure with multiscale clustering.

Corresponding node metrics can also be synthesized by sampling from an appropriate distribution. For freedom and fluidity, we also used power law distributions with representative scaling factors and exponents. The network in Figure \ref{fig:teaser} was produced using this method.

\subsubsection{Pipeline Extension Discussion}

In our pipeline, replacing Louvain optimization with our hierarchical extension of Leiden has allowed us to perform multiscale workgroup mapping in a resolution-limit--free way, adding an additional dimension to our automated analysis and creating new avenues for insights.

Future directions include developing privacy-preserving mechanisms for communicating the composition of workgroups without explicitly identifying their members. Providing a high-level understanding of the relationships between workgroup types and metric values is important for building an accurate mental model of organizational structure and dynamics. One potential approach that would balance the needs for privacy and utility would be to describe workgroups using the most frequent attributes of their individual members (e.g., role, department, and location). Anonymization techniques such as $k$-anonymity \cite{sweeney2002k} could then be applied to these descriptions such that workgroup descriptions would only be shown in cases where there are at least $k$ similar workgroups, preventing any one such workgroup being `singled out'. This approach fits well within the established privacy framework for the WPA product where protecting individual information is foundational to how the product operates. Another direction for privacy-preserving pipeline extensions would be to adapt our network synthesis techniques to use differentially-private sampling of sensitive statistics (e.g., as in \cite{proserpio2012workflow}), constructing a statistically-similar network that retains overall community structure but not potentially-identifying substructures. Such an approach could be instrumental in enabling data sharing, academic research, and 'what-if' scenario modeling and planning around organizational activities such as divestments, acquisitions, and restructuring.

\section{Conclusion}


We have proposed \emph{workgroup mapping} as way to empower organizational leaders to view the informal structure of their organizations, identify and reason about opportunities for change, and evaluate progress towards the kinds of collaboration culture that can best support business needs. Our approach is applicable to all organizations and industries where people collaborate in a digitally-mediated way, and reveals the structure and dynamics of collaboration through the visualization of two new \emph{workgroup metrics} -- the degree of change in collaborative relationships over time (`fluidity') and the extent to which those relationships span the organizational hierarchy (`freedom'). Both of these qualities are critical for understanding the informal organization of work, but neither is captured by existing methods. We have applied and evaluated these metrics in a workgroup mapping pipeline that automates the end-to-end process of network analysis and narrative visualization, supporting hundreds of presentations to organizational leaders.

The ensemble nature of our pipeline and generality of its components also means that they may be used to support diverse use cases beyond organizational research. We organize them below according to four focus areas of the visual analytics research agenda \cite{thomas2006visual}:

\emph{Analytical reasoning techniques}. Our fluidity metric was developed in the context of workgroup collaboration patterns, but could be applied to any network and level of aggregation (e.g., node, community, or full network) where change in connectivity is meaningful (e.g., to detect trends or anomalies in a dynamic network). Similarly, our freedom metric can be used to measure the alignment between any defined subset of nodes (e.g., defined by attributes not limited to community) and a corresponding tree structure (e.g., a taxonomy or typology).

\emph{Data representations and transformations}. Our hierarchical application and optimization of the Leiden algorithm can be used to extract a hierarchical community structure from any network, while our network synthesis model can be used to generate networks from any hierarchy.

\emph{Visual representations and interaction techniques}. Our approach to people network layout applies beyond the use of email exchange as a proxy for collaboration, and beyond collaboration as the basis for edge definition. Our experience shows that it can be applied to most networks induced from human behavior (e.g., web search \cite{makingsenseofsearch}).

\emph{Production, presentation, and dissemination}. Our SlideStamper service could be used for the template-based generation of any slide presentation, not limited to those based on organizational analytics (or even data graphics). Likewise, Thematic allows for harmonious styling of any interface, not just slide presentations and network visualizations. 

Ongoing work includes the integration of meetings, calls, and instant messages as additional collaboration signals, as well as the development of an interactive, privacy-preserving tool for workgroup mapping. The main area for future research, however, is to work with organizations on a longitudinal basis to understand how our metrics vary over time in response to evolving organizational forces. This include internal programs that encourage new kinds of collaboration as well as external shocks that demand new and creative ways of working together.

At the time of writing, an urgent example of such shocks is the unfolding COVID-19 crisis, where the combination of stay-at-home mandates, staff shortages, and business uncertainty threaten the survival of organizations across industries and around the world. For some organizations, the mass transition to home working also means that digital collaboration temporarily provides a complete picture of all collaboration. This has created an unprecedented opportunity to learn how properties of collaboration networks both enable and indicate effective adaptation in times of crisis -- a quality known as \emph{organizational resilience} \cite{vogus2007organizational}. Using the flexibility inherent in our pipeline, we will continue to develop and evaluate a range of metrics, including freedom and fluidity, that may help to reveal how such resilience manifests in the structure and dynamics of collaboration. Only with such insights can we guide the cultivation of more resilient organizations.


\bibliographystyle{abbrv-doi}

\bibliography{template}
\end{document}